# Technical improvements and performances of SpIOMM: an imaging Fourier transform spectrometer for astronomy

Anne-Pier Bernier[*a], Maxime Charlebois[a], Laurent Drissen[a], Frédéric Grandmont[b]

[a]Dépt. de physique, de genie physique et d'optique, Université Laval, Québec, Qc, Canada G1K 7P4
and Centre de recherche en astrophysique du Québec (CRAQ)
[b]ABB Bomem Inc., 585 Boul. Charest est, Suite 300, Québec, Qc, Canada G1K 9H4

**ABSTRACT**

We present the most recent technical improvements on SpIOMM, an Imaging Fourier Transform Spectrometer (IFTS) attached to the 1.6 telescope of the Mont Mégantic Observatory. The recent development of SpIOMM demonstrates that the concept of IFTS for ground telescopes is a promising astronomical 3D spectroscopy technique for multi-object spectroscopy and multi-band imaging. SpIOMM has been developed through a collaboration between Université Laval and the industry (ABB Bomem). It is designed for optical observations from the near UV (350 nm) to the near IR (850 nm) with variable spectral resolution. The circular FOV of the instrument covers 12' in diameter. We have recently improved the servo system algorithm which now controls the mirror displacement and alignment at a rate of ~7kHz. Hardware improvements to the servo and the metrology system will be described along with their impacts on performance in the laboratory and in observing conditions. The instrument has successfully been operated at the 1.6 meter telescope this year using the revised control systems and acquired several datacubes. We will discuss some issues regarding the sensitivity to environmental conditions implied by the use of such an instrument. An overview of the datacube reduction procedure will show some solutions proposed for observational problems encountered that affect the quality of the data such as sky transmission variations, wind, changing gravity vector and temperature.

## 1. INTRODUCTION

A major technical advance in astronomy, which combines imaging and spectroscopy led to the development of powerful instruments such as multi-object spectrographs (MOS) and integral field units (IFU). These instruments still have some limitations on spatial or spectral coverage, including small fields of view (for IFU). We propose the concept of an Imaging Fourier Transform Spectrometer (IFTS) to overcome these disadvantages. Many configurations of the Fourier Transform Spectrometer (FTS) are derivatives of the classical Michelson interferometer and have been used successfully in spectroscopy for decades. The combination of the imaging to the interferometer is a relatively new approach for a real three-dimensional device for astronomy. Its interest resides in a full FOV coverage with adequate spatial resolution, a considerably large spectral coverage and an advantageous throughput. It also offers flexibility concerning the spectral range and resolution. A first model of IFTS used in astronomy in the infrared band was operational in the early 90's[1,2] at the Canada France Hawaii Telescope (CFHT). This instrument, BEAR, produced three-dimensional data of high resolution over extended objects like planetary nebulae and galaxy nuclei[3]. It has shown the potential of IFTS as a valuable approach to 3-D spectroscopy in astronomy. To pursue the demonstration of the advantage of IFTS in astronomy, we have developed an IFTS prototype at Université Laval in collaboration with the company ABB for the 1.6m telescope at the Observatoire du Mont Mégantic. This ground-based instrument, called SpIOMM (as for "Spectromètre Imageur de l'Observatoire du Mont Mégantic") was assembled in January 2004 saw first light at the telescope later that year.

---

[*] anne-pier.bernier.1@ulaval.ca; phone 1 418 656-2131 x6192, x6193; fax 1 418 656-2040; Université Laval, Québec, QC, Canada, G1K 7P4

We have presented in previous articles[4,5] the design of SpIOMM and its core, based on the Michelson interferometer with a 30 degrees incidence angle on the beamsplitter-compensator plate assembly to minimize the latter circular size. The use of two plane mirrors reduces the number of optical surfaces encountered by the science beam from the telescope output to the camera and therefore grants a better throughput. Its optical configuration leads to a circular FOV of 12 arcmin in diameter and to a spatial resolution on the CCD detector (1300 x 1340 pixels) of about 0.55 X 0.55 arcsec per pixel. SpIOMM's optics covers the visible band from 350nm to 850nm. Retrieving spectral information from the interferometer's images implies to scan the Optical Path Difference (OPD) executed by the moving mirror. SpIOMM uses a well-known technique in Fourier spectroscopy that consist in step-by-step scanning of the path difference. A metrology combined with a dynamic alignment system guarantees the stability of the instrument. It ensures optimal modulation efficiency and precise scanning during the whole data acquisition, which is crucial to obtain the spectral information. In general, such FTS are highly sensitive on the mirror misalignment since it degrades the modulation efficiency. The details about the metrology and the servo system will be reminded in the section 3. Before introducing the old and new metrology methods, we will describe the problems experienced before the recent improvements.

## 2. OPERATIONAL ISSUES

During those years of operation we have targeted a list of issues, mostly experienced at the telescope, which mainly required improving SpIOMM's servo system.

First, the instrument has shown a sensibility to temperature and air perturbation. Temperature gradients and turbulence locally change the refraction indices of the environment in the science or metrology optical paths and leads to wavefront phase errors. We have observed strong variations especially for the metrology beam resulting in unstable metrology signals. Therefore these optical paths and the whole instrument need to be sealed to avoid any turbulence within it. Also, we have observed that the temperature fluctuations affect the mirror alignment values although the instrument structure is athermal. This variation is due to thermal expansion of a smaller structure, namely the three screws that anchor the fixed mirror to the interferometer module. The slow temperature variations during the night (between 5 and 10 °C) differently affect the stiffness on those three pressure points on the mirror and therefore induce a slow drift of the alignment angles. Consequently, we have to assure that the fixed mirror is aligned at the mean temperature such as the alignment values are set in the center of the dynamic range possible for the moving mirror. So far, this solution has been sufficient to operate for a period as long than a season or a change of less than 15 °C.

Secondly, we have found that the instrument is affected by gravity. We have re-assembled the whole instrument structure in 2006 to assure a better frame rigidity. It did improve the structure stiffness but the alignment values still depend on the inclination of the telescope (as well as temperature like discussed above), and thus of the instrument. To assure a sufficient angle correction range, the alignment adjustment is done at the zenith that corresponds to the position of equilibrium where the moving mirror is displaced on the vertical axis. In this manner, at different telescope positions the gravity induces angle variations that lie within the dynamic alignment range.

Thirdly, we have noticed that the instrument is very sensitive to vibrations. Any vibration affects the interferometer mirror alignment and causes temporary and random losses in modulation efficiency. The first servo system was conceived such as to correct small amplitude and relatively sudden misalignments of interferometer's moving mirror at each step due to its mechanical displacement or from external vibrations caused by the wind on the telescope structure in the dome. Additional vibrations of higher amplitude and frequency can overload the metrology and servo system. One of the first causes of additional vibration was introduced by the degradation of the inchworm actuator in charge of the displacement of the moving mirror. It could periodically produce jolts during its course that could make the servo system loose its phase reference (like a glitch). This produces a discontinuity in the interferogram signal during the acquisition and therefore results in computational errors of the associated spectrum. Also, we had no evidence where in the acquisition sequence the damaging glitches would occur. The actuator has been repaired since then but further improvements in metrology and dynamic alignment were considered to overcome such possible mechanical degradation. Simply the increase of the speed of the servo program can guarantee a good sampling and a detection of the phase wrapping during a glitch of the actuator.

Another cause of vibration has appeared later during the first year of operation and was due to a modification of the air pump that feeds the telescope primary mirror's support system. This modification brought more important vibrations on the instrument. These periodic vibrations from the pump are affecting mostly the detection of the metrology signals. Consequently it induces high variability and uncertainties in the calculation of the correction angles to be applied by the dynamic alignment system. We first found a quick solution for it by placing the new pump outside the telescope and conveying the air jet with a hose to the primary mirror. For a long-term solution, we considered that the improvement of the metrology system would decrease its susceptibility to such vibrations.

Finally, the problem that has convinced us to change the metrology and the servo system is caused by a bias in the metrology system concerning the fringe pattern detection. We have demonstrated that the old method leads to correction values biased because of the calculation from an IR fringe pattern (containing around 4-5 fringes) produced by intentional misalignment of the metrology beam only. This bias induced a periodic variation of the alignment correction as the interferometer mirror was completing its scan. Consequently it caused a periodic loss of modulation efficiency during the acquisition that produces artefacts in the resulting spectra. Hence it was imperative to change the concept of the metrology system to avoid such problems and also to thwart the vibrations experienced at the telescope.

Recently, we have developed a new method to measure the mirror misalignment by detecting a metrology flat fringe instead of a metrology multiple fringes pattern. This new metrology technique assures a more stable and rapid control of the mirror. Its design will be presented in the next section and its performances will be compared to the previous system in the section 4. We have verified that the improvements on the metrology and servo system yield a complete solution for most of the problems listed previously. Since June 2007 the instrument has achieved more than 30 complete datacubes of astronomical objects[8] and has proven its reliability in the operating environment.

## 3. MODIFICATION OF THE METROLOGY SYSTEM CONFIGURATION

Our principal goals for improving the instrument are to speed up and stabilize the servo response. In order to explain the required modification performed on the metrology system, we first summarize the principle of the old method. The position and alignment of the moving mirror were detected by the metrology system and were acquired by a servo-computer at a rate of 2kHz. This feedback is given by the interference pattern of an infrared 1550nm laser beam expanded to a diameter of almost 1 inch. This metrology source is directed through the interferometer and is located in the telescope' primary mirror obscuration circle on the interferometer mirrors. Thus it is operating in parallel with the science beam. A fringe pattern of this IR source is created by a wedge on the fixed mirror at the interferometer output and is detected orthogonally by two 32 pixels detector arrays. Therefore those detectors record a sinusoidal type signal with a changing frequency. Figure 1 shows the fringe pattern behavior as the mirror is misaligned. The frequency of these metrology signals changes with the inclination of the mirror. The servo code was calculating by Fast Fourier Transform (FFT) the frequency of the detected fringe pattern. As the mirror is moving, the fringes are swept on the detector keeping track of the relative phase of the corresponding OPD. Consequently, the alignment and the positioning correction of the mirror are calibrated in terms of, respectively, the frequency and the phase of the recorded infrared fringes. Once the correction values are calculated, the servo computer sent back the proper correction voltages to the dynamic alignment piezo which then realign the moving mirror.

The moving mirror is mounted on a parallelogram porch swing-type mechanism which offers frictionless displacement[6]. It is moved by a piezo-based stepper motor mounted in series with a small range high frequency response piezo (called OPD piezo). The stepper actuator can travel a Maximum Path Difference (MPD) of 2 cm. The ensemble allows a high stiffness (10 N/μm) to ensure good passive stability of the OPD. A dynamic alignment system using piezoelectric (DA piezo) combined with the mirror displacement mechanism is required for a precise positioning to evenly sample the interferograms. Also, this maintains the mirror position and alignment during the exposure time subject to gravity, vibration and thermal perturbation.

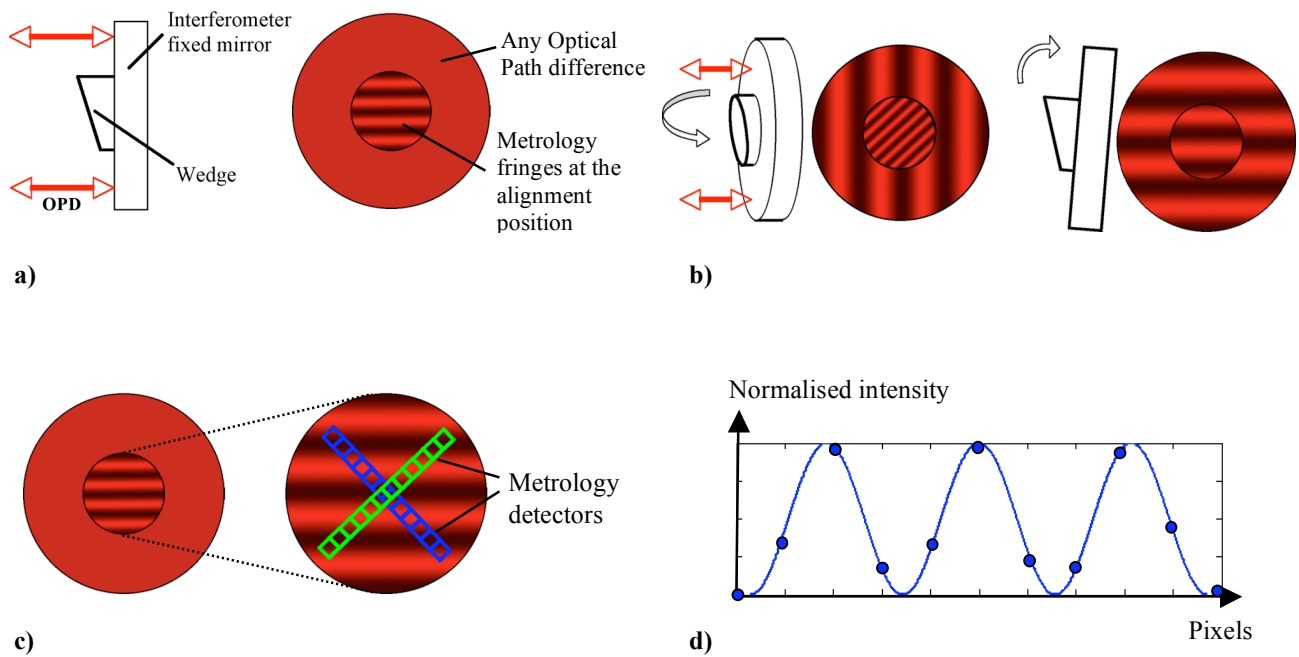

**Figure 1** : Design of the old metrology system consisting in the detection of the fringe pattern from an IR source created by a wedge placed at the center of the fixed mirror (a). The misalignment angle in the 2 orthogonal axes will change the frequency of the fringe pattern (b) and its signal (d) is recorded by 2 detectors orthogonally placed at the interferometer's output (c).

As we have mentioned before, this old metrology method had an intrinsic variability in the frequency value detected as a function of the OPD for a constant alignment. Therefore it caused a periodic bias of the alignment correction. We concluded that this bias could be avoided if the mirror alignment is measured by a flat fringe (uniform) pattern directly corresponding to the correct alignment instead of a multiple fringe pattern based on an intentional misalignment (wedged mirror). In order to detect an angle and a path difference (phase) on metrology flat fringe we added a simple wave-retarder in the metrology beam as shown in figure 2. The wave-retarder consists of a circular layer of glass at the center of the fixed mirror. Its thickness makes a retardation of $\lambda/8$ at the metrology source wavelength (1550nm) so that after the reflection on the mirror, the metrology beam is retarded by $\lambda/4$ since it passes twice in the layer. The diameter of the wave-retarder layer is smaller than that of the metrology beam. Therefore its detector arrays distinguish a circular inner part of the beam that is retarded by $\pi/2$ compared to the outer annulus part which has not passed through the retardation layer.

Figure 2 (c) shows the flat fringe pattern as seen by the metrology detector. We can clearly identify the inner part of the metrology beam that passes into the wave-retarder region compared to the signal extremities outside of this region of retardation.

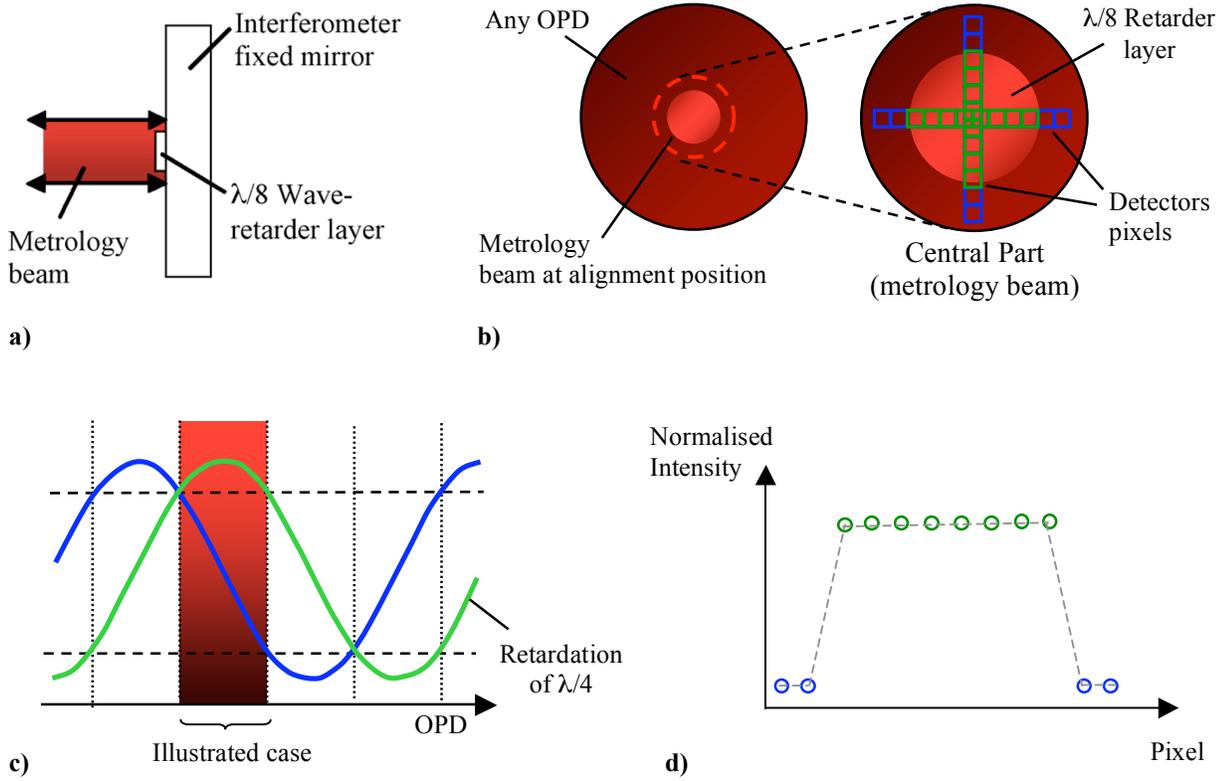

**Figure 2**: Design of the new metrology system consisting in the detection of a flat fringe pattern (of the IR source) comprising a smaller circular region of λ/4 retardation created by a wave-retarder layer placed at the center of the fixed mirror (a). The flat fringe pattern is recorded by the same 2 detectors orthogonally placed at interferometer output (b). The signal of a pixel located in the inner region of retardation of the metrology beam will be delayed by π/2 compared to the signal of a pixel outside (c). For a specific OPD, (d) shows the shape of the aligned intensity pattern read by one of the detectors.

The following equations (1 and 2) present the relation between the intensity pattern I(y) of the metrology beam (outside and inside the retardation region respectively) as a function of the misalignment angle α and the OPD Δz (correction parameters). The equations for the x axis are identical.

$$I_{ext}(y) = \cos\left(\frac{2\pi}{\lambda}(\Delta z + 2\alpha y)\right) = \sin\left(\frac{2\pi}{\lambda}(\Delta z + 2\alpha y) + \frac{\pi}{2}\right) \quad (1)$$

$$I_{int}(y) = \cos\left(\frac{2\pi}{\lambda}(\Delta z + 2\alpha y) - \frac{\pi}{2}\right) = \sin\left(\frac{2\pi}{\lambda}(\Delta z + 2\alpha y)\right) \quad (2)$$

Therefore, we can isolate the misalignment angle α from the two previous relations such that:

$$|\Delta z + 2\alpha y| = \frac{\lambda}{2\pi}\left(\arccos(I_{ext})\right) = \frac{\lambda}{2\pi}\left(\arccos(I_{int}) + \frac{\pi}{2}\right) \quad (3)$$

This expression shows that small values of α and the OPD Δz (relative phase) can be determined by a linear regression on most of the pixels of the appropriate region (interior or exterior) chosen. For a fixed misalignment angle, the intensity patterns vary with the OPD and the servo code simultaneously evaluates which region has the best intensity pattern for a linear regression. This method is advantageous for its rapid calculation of the correction values compared to the previous one that implied a numerical calculation of the Fourier transform of the intensity patterns. Moreover, such an intensity pattern read by the detector pixels behaves steadily and does not suffer from the variability caused by additional vibrations.

## 4. METROLOGY AND SERVO SYSTEM PERFORMANCE

The old servo system was calculating and correcting the alignment at a rate of 2.5 kHz. Since the new technique does not use numerical computation of the Fourier transform of the metrology signal, its servo speed increased at a rate of 7.8kHz. This increase by almost a factor of three is considerable for keeping track of high frequency vibrations such as the jolts (glitches) of the stepper actuator. Figure 3 shows the OPD detected by the servo system in units of number of wave fringes at 1550nm during a displacement done by the stepper actuator. We can see that one jolt happened during this displacement. The zoom on the glitch clearly shows a sufficient and continuous sampling of the vibration amplitude. Therefore the speediness of the new servo system assures no break of the metrology feedback and a more stable alignment correction against numerous vibrations.

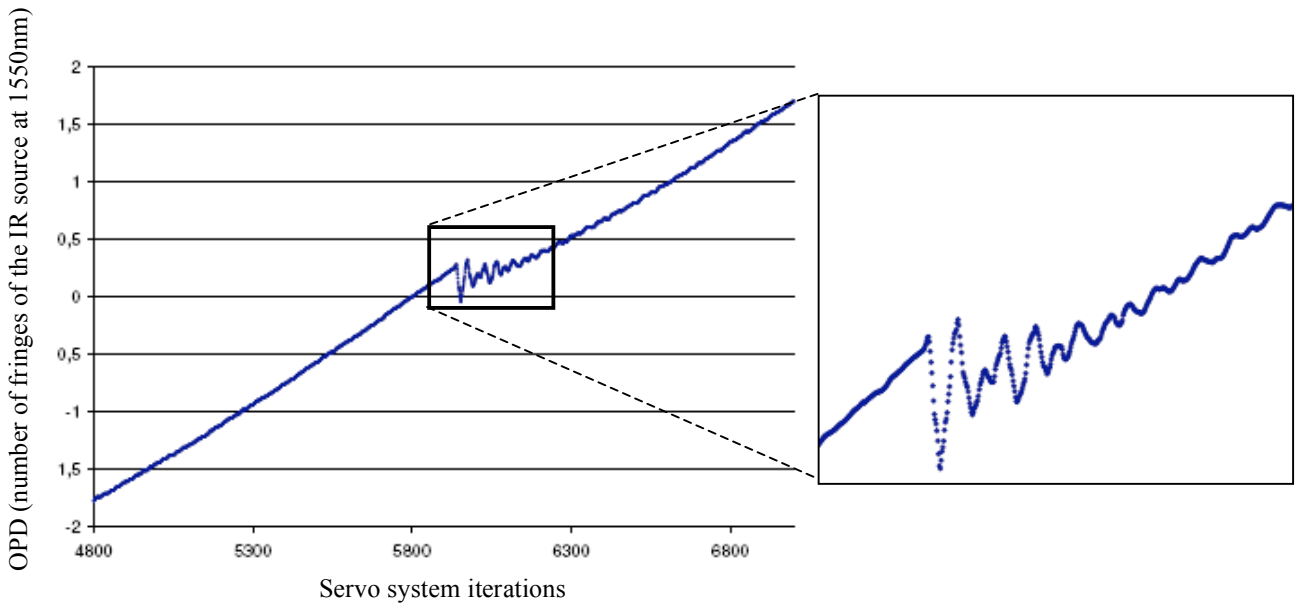

**Figure 3**: Measurement of the stepper actuator's displacement with a jolt in terms of the OPD (number of fringes of the 1550nm laser source) as a function of the iterations for the new servo code at a rate of 7800 iterations per second.

We have observed with time that the metrology signal intensity can change. The maximum intensity detectable for each pixel can vary because of dust or optical surface anomalies in the metrology beam optical path and therefore can affect the stability of the metrology signal. Before we discuss more on the stability of the metrology signal (for the flat fringe pattern) for comparison with the old method, we must introduce an additional issue concerning an optical defect in the interferometer module. We have observed optical ghosts in our images of the IR metrology beam at the interferometer exit, which are created by reflections on each optical surface of the beamsplitter-compensator assembly. Figure 4 illustrates these reflections, which contaminate the metrology beam itself. We added a mask at the exit of the metrology laser to shape the beam in a form of a cross. This cross shape of the beam minimizes the impact of the overlaying ghosts. But there is still a ghost beam that is reflected close to the main beam and contaminates one side the arms of the cross as shown on figure 4. Therefore, we had to consider that the metrology pixel signals affected by the ghost beam had to be rejected from the servo calculations because of their highly noisy interference. For the old metrology method, this was devastating on the multiple fringes pattern since it creates a discontinuity of the sinusoidal signal, making it harder to calculate the frequency of the multiple fringes pattern. This was probably participating to the strong variability of the alignment values detected by the old metrology system.

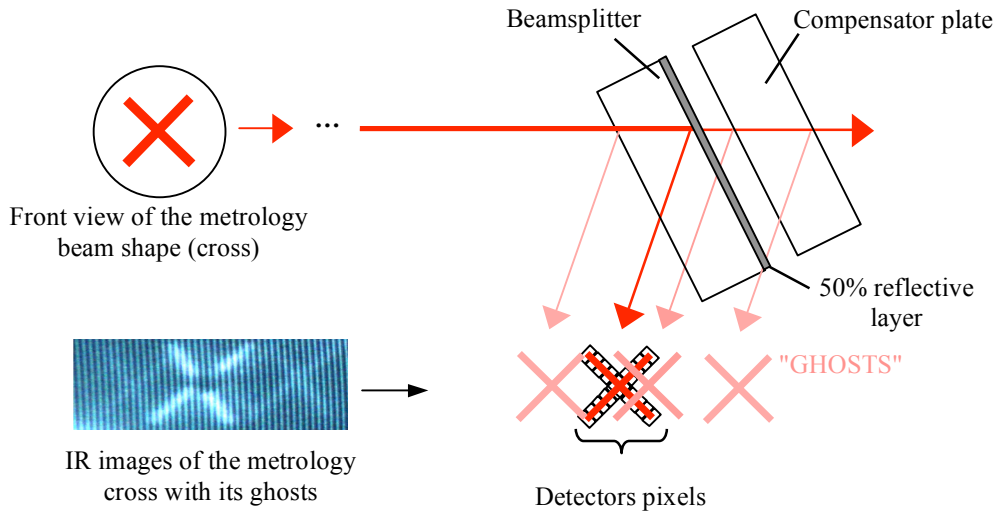

**Figure 4** : Simplified schema of the metrology optical path through the interferometer that shows reflections creating the ghost beams.

In the case of the new metrology method, we still reject the pixels affected by the ghosts. But the resulting signal which consists of a flat fringe pattern is less affected by this loss for the servo calculation. Figure 5 presents the metrology signals of each detector (x and y axes) obtained during 4 different days to show the variability of each pixel. It also distinguishes the best pixels on each detector array to consider in the servo calculation. The pixels located in the ghost region, at the border of the wave-retarder and at the ends of the pixel arrays vary in time. Consequently we don't consider those pixels and the servo system performs well even without those cuts since the linear regression needs less point in the signal than does the Fourier transform of the old metrology pattern.

We have performed a datacube of a monochromatic flat field (with a HeNe laser) in order to verify the constancy of the new servo correction during a whole acquisition. Good and constant modulation efficiency at each step certifies that the moving mirror is well aligned during the whole acquisition. The modulation efficiency is determined by the contrast between the dark and bright fringes of the monochromatic source at each step. We compare the performance of the old and new servo system in Figure 6. It shows the measured modulation efficiency (in %) at each step of the datacube. The middle region of the acquisition sequence corresponds to the zero path difference (ZPD) where the number of fringes in the field of view decreases to one and therefore the calculation of the modulation efficiency is incorrect. So we don't consider the results in this region. Mostly we evaluate the mean efficiency at the beginning and at the end of the sequence and its variations.

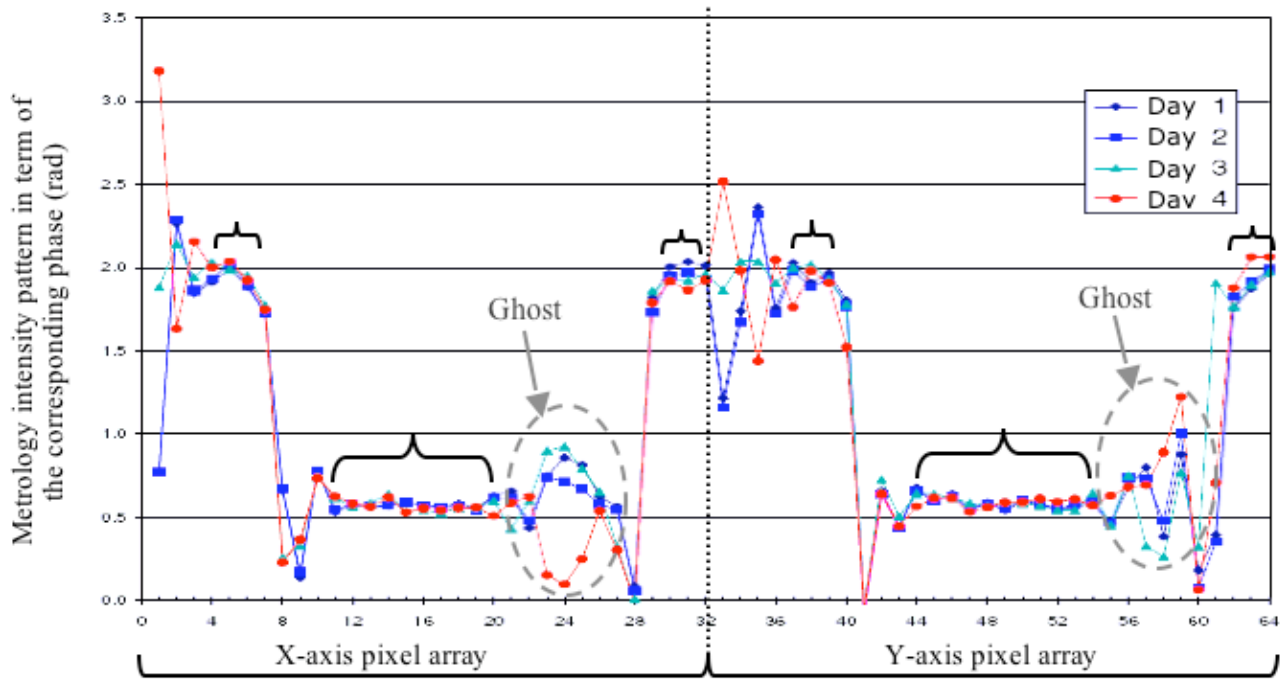

**Figure 5**: Intensity pattern of the new metrology method recorded during 4 different days and the pixels chosen for their stability.

From figure 6, we evaluate a mean modulation efficiency of 74% with a deviation of 8% for the datacube obtained with the old system. Moreover, its variability is periodic. As we have mentioned previously, this behavior produces artifacts in the analyzed spectrum since it adds a frequency in the interferogram. Therefore a random and small deviation is preferred since it will add a white noise (uniformly distributed) in the spectrum.

The new servo system gives a better mean modulation efficiency of 78% with a deviation of 2%. Its deviation varies randomly and is 4 times smaller than with the old system. This new metrology technique definitely shows an obvious gain in its capability. It has greatly improved its performance and has accomplished our objectives, which are a more stable and a faster servo response.

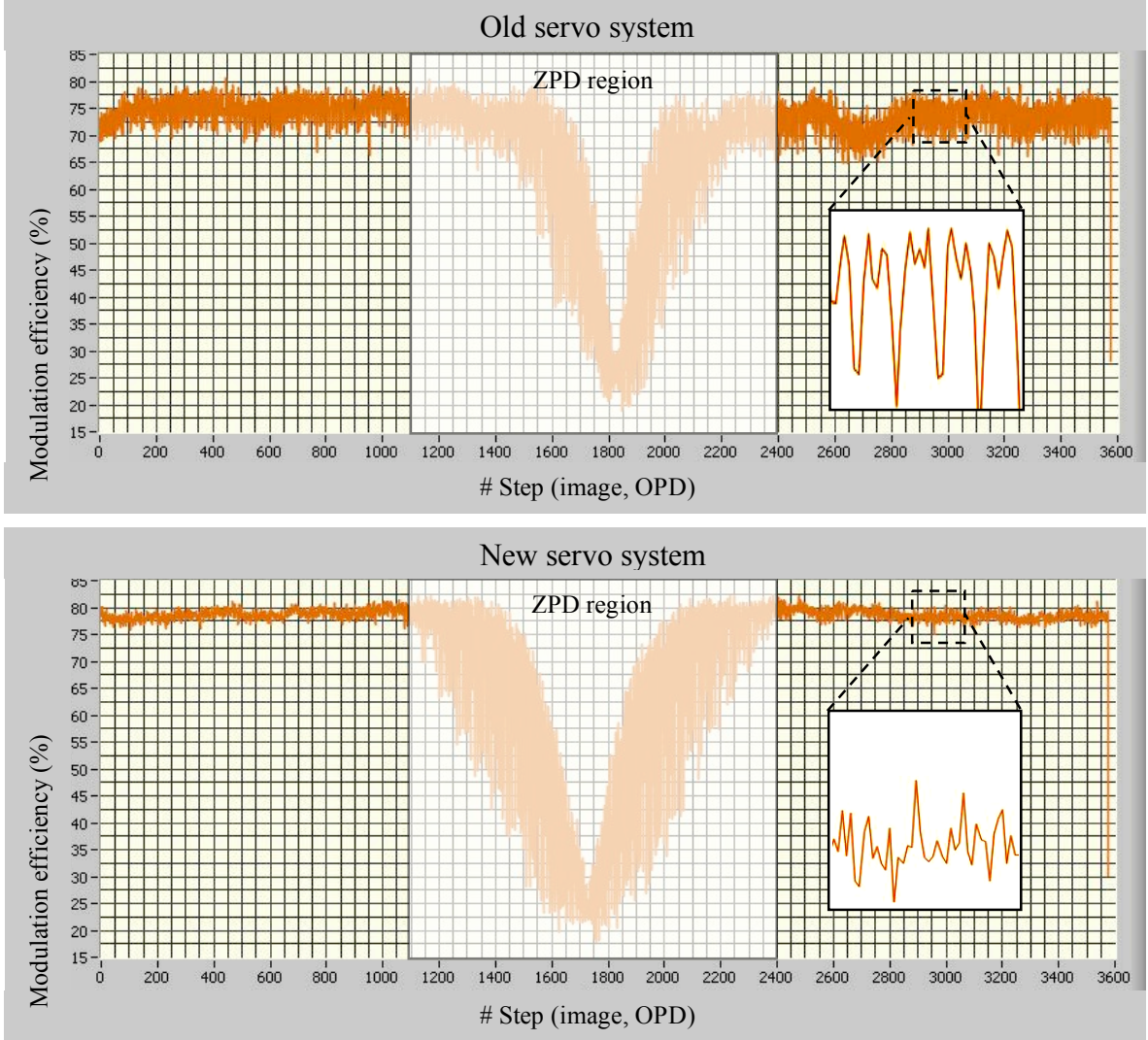

**Figure 6** : Modulation efficiency obtained from datacubes of a monochromatic (HeNe laser, 632nm) flat field with the old and the new servo system.

## 5. DATA REDUCTION

The IFTS datacube consists in a series of images corresponding to the steps scanned by the interferometer mirror. Globally, the size and the number of the steps fix the spectral band to be sampled and the total OPD traveled by the mirror, proportional to the spectral resolution. These are preliminary acquisition parameters that need to be set before the scan. The intensity of a given pixel in a raw datacube as a function of the step number is an interferogram of the light source as a function of the OPD. Our data reduction takes care of both the image and interferogram dimensions, which are necessary to obtain the hyperspectral cube. The principal steps of the data reduction are listed in the next subsections. All of the following data manipulation has been developed in the IDL platform and most could also be executed by IRAF tasks.

## 5.1 Image calibration

First, typical image calibration is applied on each of the datacube images such as the CCD camera bias removal and flat field frame correction. The purpose is first to remove artifacts like the glow from camera noise (electronics, hot pixels, etc). The SpIOMM FOV is circular such that there is vignetting in the images corners. Thus flat-fielding correction re-adjusts the light level of the image corners. Also, it ensures as far as possible a uniform image with dark and light patches (caused for example by dust spots and by the different sensitivity of the pixel) eliminated.

## 5.2 Photometric alignment of the images in the cube

The alignment of the images in a datacube is very important in order to analyze the signal from a specific pixel. During a typical datacube acquisistion, usually more than an hour, the FOV can slowly drift because of flexion in the instrument. We therefore need to realign each frame to correctly compute the spectrum of every light source in the field. We therefore select a dozen bright, but not saturated, stars in the first frame, which is used as a reference and compute their centroid in each image of the cube. The average centroid difference is then used to realign each image of the cube. Typical translations during a large datacube of a galaxy amount to less than 3 arcseconds. Our images never have suffered from rotation nor stretching, so there is no need for such a correction.

## 5.3 Correction of the sky background

Before we transform the interferograms into spectra, we want to subtract the modulation from sky background. This partly eliminates the spectral contribution from the sky background. We simply pre-select sky regions devoid of stars or nebulae and determine the median value of sky background intensity and subtract this value for each image.

## 5.4 Correction for the sky transmission

As mentioned before, the period of datacube acquisistion is usually more than an hour and varies depending of the datacube parameters and the integration time of the object. The sky transmission sometimes fluctuates over that period and affects highly the signal of the interferogram; this is due to the passage of thin clouds or simply the change of airmass. It can add low and high frequency contributions to the spectra. To avoid such impact we want to normalize the baseline signal of the interferogram. To do so, we use the intensity variation of the stars in the FOV. The interferogram signal of polychromatic source modulates only at the ZPD. On either side of the ZPD, the signal intensity is supposed to be flat unless the sky transmission fluctuates (clouds, airmass, etc). Therefore we determine the mean sky transmission signal from the interferograms of a list of stars. We also evaluate the ZPD region in the stars interferogram and we interpolate a line in that part of the signal. This particular detail demand a very stable measurement when passing the ZPD part during the acquisition at the telescope. Besides we can pause and recapture previous steps during the acquisition when some cloud appears in the FOV. The resulting sky transmission signal is then used for normalization of the interferogram signal.

## 5.5 Correction for cosmic rays and satellite traces

A cosmic ray (or a satellite trace) in a frame of the datacube will add a very high intensity and sharp peak at the corresponding pixel interferogram. The cosmic ray intensity has the form of the Dirac function that is convolved with the interferogram signal[7]. We know that the Fourier transform of a normalized and shifted Dirac function yields a sinus in the frequency domain. Therefore, the impact in the spectrum after the transformation of such interferogram is an important periodic noise addition, reducing the real spectral lines level. Thus we want to reject the cosmic ray not only for imaging reason, but also for spectroscopy reason. We have developed an IDL program to detect the pixels affected by cosmic rays and the satellite trace. The detection program uses the interferogram signal and evaluate when the pixels have more than $3\sigma$ of signal. The intensity of the detected pixel is replaced by the mean intensity value of the neighbor's pixels of the affected image to minimize the noise.

## 5.6 Datacube transformation

Before transforming the datacube into a cube of spectra we must apply some operations on the interferogram signal. First, the pre-processed interferogram should have a null mean in order to have its extremities values approaching zero. Therefore, we remove the mean value to each interferogram. We then multiply the interferogram with an apodization function in order to minimize the side lobes in the spectral lines created by the truncation of the signal (finite number of points)[4]. We have tested some apodization windows with our data to find the function that minimizes the side lobes in the spectrum. The gaussian function provides the best results. Finally, a zero-filling operation is performed on the interferogram in order to increase the number of points up to the next higher power of 2. This is useful for further spectral interpolation and for a faster processing discrete Fourier transforms[7]. We then apply the discrete Fourier transform to each processed interferogram of the datacube (up to 1.7 million) in order to obtain a datacube of spectra. In parallel we compute the spectral axis scale from the datacube sampling parameters and we interpolate the total number of points on a graduation in nanometers. This rescaling and interpolating operation takes also in account a calibration dataset that corrects for off-axis contributions on the whole FOV.

## 6. CONCLUSION

In operational situation, the instrument has proven again its advancement. In 2007 and 2008, SpIOMM has acquired over 30 datacubes of extended objects, principally nebula and galaxies. At the telescope, the servo system has not been sensible to the primary mirror pump's vibration as it was before. Nevertheless, it can still be affected by vibrations of high amplitude, which are not usual anyway. The actuator's glitches are no longer a problem with the servo speed of 7800 iterations per seconds and we have not experienced OPD (phase) loss during acquisition since the improvement. A complementary article (Drissen et al., these proceedings)[8] presents some of the datacubes obtained during the last year and shows the numerous possibilities of SpIOMM for the integral spectroscopic mapping of extended astrophysical objects.

## ACKNOWLEDGMENTS

We would like to thank Jean-Pierre Maillard, who has pioneered the use of FTS at the Canada-France-Hawaii telescope, for his continuous support since the early stages of the development of SpIOMM, as well as ABB Bomem, for its technical support. SpIOMM was funded by the Canadian Foundation for Innovation, the Canadian Space Agency, Québec's FQRNT, Canada's NSERC, and Université Laval. Finally, we would like to thank NSERC, ABB Bomem and Canada's MITACS for the Industrial Postgraduate Scholarship.